\documentclass[journal]{IEEEtran}
\usepackage{epsfig,amsfonts,amsbsy,bm,mathrsfs}
\usepackage{mathrsfs} 
\usepackage{graphicx,cite,amssymb,amsmath,bm}
\usepackage{mathtools}
\usepackage{amsmath} 
\usepackage{etex}
\usepackage{amsmath,amssymb}
\usepackage{lipsum}
\usepackage{array,verbatim}
\usepackage{makecell}
\usepackage{siunitx}
\usepackage{float}
\usepackage{xcolor}
\usepackage{colortbl}
\usepackage{tikz}
\usepackage{ctable}
\usepackage[english]{babel}
\usepackage{color}
\usepackage[utf8]{inputenc}
\usepackage[T1]{fontenc}
\usepackage{balance}
\usepackage{balance}
\usepackage{perpage}
\usepackage{algorithm}
\usepackage[noend]{algpseudocode}
\usepackage{subcaption}
\usepackage{graphbox,graphicx}
\usepackage{xcolor}

\addto\captionsenglish{}
\makeatletter
\def\algbackskip{\hskip-\ALG@thistlm}
\makeatother

\newcommand{\mep}{\mathtt{x}}
\newcommand{\row}{\mathsf{r}}
\newcommand{\col}{\mathsf{c}}
\newcommand{\rmT}{\mathsf{T}}
\newcommand{\Q}{\mathsf{Q}}
\newcommand{\Bmat}{\mathbf{B}}
\newcommand{\Cmat}{\boldsymbol{C}}

\renewcommand{\a}{\boldsymbol{a}}
\renewcommand{\b}{\boldsymbol{b}}

\newcommand{\nc}{n}

\newcommand{\kc}{k}
\newcommand{\dc}{d_{\mathsf{c}}}

\newcommand{\lr}{\boldsymbol{r}}

\newcommand{\lc}{\boldsymbol{c}}

\newcommand{\rr}{\boldsymbol{R}}
\newcommand{\rbol}{\boldsymbol{r}}

\newcommand{\cc}{\boldsymbol{C}}

\newcommand{\dmin}{d_\mathsf{min}}

\newcommand{\lalone}{\boldsymbol{L}}

\newcommand{\w}{w}
\newcommand{\ww}{\boldsymbol{w}}
\newcommand{\BB}{\mathsf{B}}

\newcommand{\dv}{d_{\mathsf v}}

\definecolor{mygreen}{rgb}{0.0, 0.45, 0.0}

\def\forcemath#1{\ifmmode #1 \else $#1$\fi}

\newcommand{\Pue}{P^{\mathsf{e}}}
\newcommand{\fue}{f^{\mathsf{e}}}
\newcommand{\Que}{Q^{\mathsf{e}}}
\newcommand{\Puc}{P^{\mathsf{c}}}
\newcommand{\Quc}{Q^{\mathsf{c}}}
\newcommand{\fuc}{f^{\mathsf{c}}}

\newcommand{\ham}{\mathsf{d}_\mathsf{H}}

\newcommand\scalemath[2]{\scalebox{#1}{\mbox{\ensuremath{\displaystyle #2}}}}   

\ifCLASSINFOpdf
\else
\fi

\let\originalleft\left
\let\originalright\right
\renewcommand{\left}{\mathopen{}\mathclose\bgroup\originalleft}
\renewcommand{\right}{\aftergroup\egroup\originalright}

\begin{document}

\title{Binary Message Passing Decoding\\ of Product-like Codes}
\author{
	Alireza Sheikh \IEEEmembership{Student Member, IEEE}, Alexandre Graell i Amat, \IEEEmembership{Senior Member, IEEE}, \\and Gianluigi Liva, \IEEEmembership{Senior Member, IEEE}
	\thanks{This paper was presented in part at the European Conference on Optical Communications (ECOC), Rome, Italy, 2018, and the International Symposium on Turbo codes $\&$ Iterative Information Processing, Hong Kong, 2018.}
	\thanks{This work was funded by the Knut and Alice Wallenberg Foundation.}
	\thanks{A. Sheikh and A. Graell i Amat are with the Department of Electrical Engineering, Chalmers University of Technology, SE-41296 Gothenburg, Sweden (email: {asheikh, alexandre.graell}@chalmers.se).}
	\thanks{G. Liva is with the Institute of Communications and
		Navigation of the German Aerospace Center (DLR), M\"unchner Strasse 20, 82234 We{\ss}ling, Germany (email: gianluigi.liva@dlr.de).}}

\maketitle

\begin{abstract}
	We propose a novel binary message passing decoding algorithm for product-like codes based on bounded distance decoding (BDD) of the component codes. The algorithm, dubbed iterative BDD with scaled reliability (iBDD-SR), exploits the channel reliabilities and is therefore soft in nature. However, the messages exchanged by the component decoders are binary (hard) messages, which significantly reduces the decoder data flow. The exchanged binary messages are obtained by combining the channel reliability with the BDD decoder output reliabilities, properly conveyed by a scaling factor applied to the BDD decisions. We perform a density evolution analysis for generalized low-density parity-check (GLDPC) code ensembles and spatially coupled GLDPC code ensembles, from which the scaling factors of the iBDD-SR for product and staircase codes, respectively, can be obtained. For the white additive Gaussian noise channel, we show performance gains up to $0.29$ dB and $0.31$ dB for product and staircase codes compared to conventional iterative BDD (iBDD) with the same decoder data flow. Furthermore, we show that iBDD-SR approaches the performance of ideal iBDD that prevents miscorrections. 
\end{abstract}

\begin{IEEEkeywords}
	Binary message passing, bounded distance decoding, complexity, hard decision decoding, product codes, staircase codes.
\end{IEEEkeywords}

\section{Introduction}

\IEEEPARstart{A}{ renewed} interest in the design of iterative coding schemes has been recently triggered by the need of efficient error correction mechanisms for very high throughput applications. Turbo and low-density parity-check (LDPC) codes  have been shown to be capable of approaching the channel capacity under belief propagation (BP) decoding \cite{Berrouetal93,Richardson2001}. However, applications requiring transmission at data rates of several hundreds of Gbps (such as optical transport systems \cite{Justesen2010}) pose a challenge to the implementation of fast BP decoders. The difficulty is especially due to the handling of the internal decoder data flow when soft messages are exchanged within the iterative decoder. In this context, attempts to reduce to complexity of soft-input soft-output (SISO) LDPC decoders have been undertaken, e.g., in \cite{Darabiha2010, Mohsenin2010, Angarita2014,Cushon2016}. An alternative line of research is to resort to hard decision decoding (HDD) for such applications. HDD reduces the decoder data flow \cite{staircase_frank} at the expense of some performance loss compared to (BP) soft decision decoding (SDD) \cite{sheikhAIR17}.

Among the coding schemes that are particularly suited for high throughput HDD, product codes (PCs) \cite{Elias1954} gained a large attention. The iterative decoding of PCs dates back to 1968 \cite{Abramson1968}, and it regained attention after employing SDD based on the Chase-Pyndiah decoding algorithm, which improved the coding gain extensively \cite{Pyndiah1998}. Several works tackled the complexity reduction of the Chase-Pyndiah decoder \cite{Dweik2009,Lu2014,Mukhtar2016,Dweik2018,Ahn2018}. However, these approaches still require the  exchange  of soft messages between component decoders, which prevents from achieving very high throughputs.  Recently, HDD  of product-like codes with bounded distance decoding (BDD) of the component codes, which we refer here to iterative BDD (iBDD), has been considered  for high-throughput optical communications due to their excellent performance-complexity tradeoff, see, e.g., \cite{Li2011,Jian2013,Sheikh2017Shap}.
For instance, PCs have been adopted by the optical submarine standard \cite{ITUT}.

To keep up with the increasing demand of coding gains,
	performance improvements of the coding scheme coupled with a reasonable decoding complexity are necessary.
	In \cite{Hag18}, an algorithm that exploits conflicts between component decoders in order to assess their reliabilities even when no channel reliability information is available, was proposed. The algorithm, named anchor decoding (AD), improves the performance of iBDD at the expense of some increase in decoding complexity.  However, the performance is still limited by the binary (i.e., hard) nature of the decoder input.
	The use of HDD is typically motivated by the need of operating with low-complexity analog to digital converters (ADCs) and  the above mentioned limitations in the internal decoder data flow. Whereas the latter motivation is hard to circumvent, the former motivation is often irrelevant for several practical applications. In this case, the use of soft information at the decoder input may be considered if the complexity of the decoding algorithms is kept close to that of HDD. Following this paradigm, one proposal is to concatenate an inner  code  for which SDD can be performed with limited complexity, with an outer HDD code  \cite{Zhang2017,Barakatain2018}. 

 In this paper, we propose a low-complexity binary message passing decoding algorithm for product-like codes that relies on BDD of the component codes. 
 The proposed algorithm exploits the channel reliabilities and thus is \emph{soft} in nature. However, the component decoders exchange only hard decisions (i.e., binary messages), which yields a decoder data flow equal to that of conventional iBDD. The binary messages are obtained by combining the BDD outputs with the channel soft information similar to the approach proposed in \cite{Lechner2012} for the decoding of LDPC codes.
The algorithm, which we dub iterative bounded distance decoding with scaled reliability (iBDD-SR), relies on the weighted sum of the BDD output with the channel log-likelihood ratio (LLR), where the BDD decoder output reliability is conveyed by a scaling factor applied to the BDD outbound messages.

This work is the extension of the conference paper \cite{She18}, where we originally proposed iBDD-SR for PCs, and where the scaling factors were derived based on simulation results. This prevented the extension of iBDD-SR to other product-like codes, as the search of the scaling factors based on simulations for product-like codes such as staircase codes proved infeasible. Here, we derive a density evolution  under iBDD-SR for generalized LDPC (GLDPC) code ensembles and spatially coupled GLDPC (SC-GLDPC) code ensembles,  based on extrinsic BDD of the component codes, that allows us to derive the scaling factors. The derived density evolution analysis rigorously takes care of miscorrections, and follows the same principles as the density evolution proposed in \cite{JianPfister2017} for the iBDD case. In particular, the density evolution for GLDPC and SC-GLDPC ensembles allows to obtain the optimal (asymptotically in the large blocklength limit) scaling factors for iBDD-SR of PCs and \emph{spatially coupled} product-like such as staircase codes, respectively.  Our simulation results show that iBDD-SR remarkably outperforms iBDD for both PCs and staircase codes, and performs very close to the (genie-aided) miscorrection-free iBDD.

The proposed algorithm is a promising solution for very high-throughput applications such as fiber-optic communications. For instance, the 400G ZR standard for transmission at 400Gbps over data center interconnect links up to 100 km, has agreed on an FEC scheme consisting of the concatenation of an inner Hamming code decoded soft and an outer staircase code with hard decision decoding.

Related work includes \cite{She18b,Yibitflip}, and \cite{She18b1}. All these algorithms require the knowledge of the least reliable bits in the decoding process, and are significantly more complex than the proposed algorithm.

\noindent \emph{Notation:} We use boldface letters to denote vectors
and matrices, e.g., $\boldsymbol{x}$  and $\boldsymbol{X}$, with $x_{i,j}$ representing the element corresponding to the $i$-th row and $j$-th column of $\boldsymbol{X}$. 
$|a|$ denotes the absolute value of $a$,
$\left\lfloor a \right\rfloor$ the largest integer smaller than or equal to $a$, and $\left\lceil a \right\rceil$ the smallest integer larger than or equal to $a$.
A Gaussian distribution with mean $\mu$ and variance $\sigma^2$ is denoted by $\mathcal{N}(\mu ,\sigma^2)$. The Hamming
distance between vectors $\a$ and $\b$ is denoted by $\ham(\a,\b)$. 

\section{Preliminaries}
\label{sys_mod} 

Our focus is on two main classes of product-like codes, namely two-dimensional PCs and staircase codes.

\subsection{Product Codes and Staircase Codes}\label{PCstaircase}

We consider two-dimensional PCs with the same component code for the row and column codes. However, we remark that the proposed decoding algorithm  extends in a straightforward manner to more general PCs, where the row and column codes are not necessarily the same.

\begin{figure}[!t]
	\includegraphics[align=c,scale=0.4]{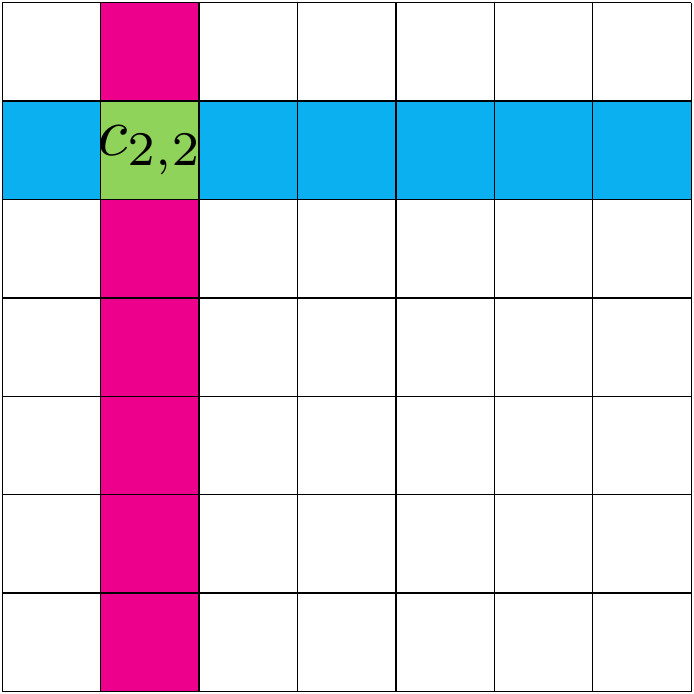}
	\includegraphics[align=c,scale=0.78]{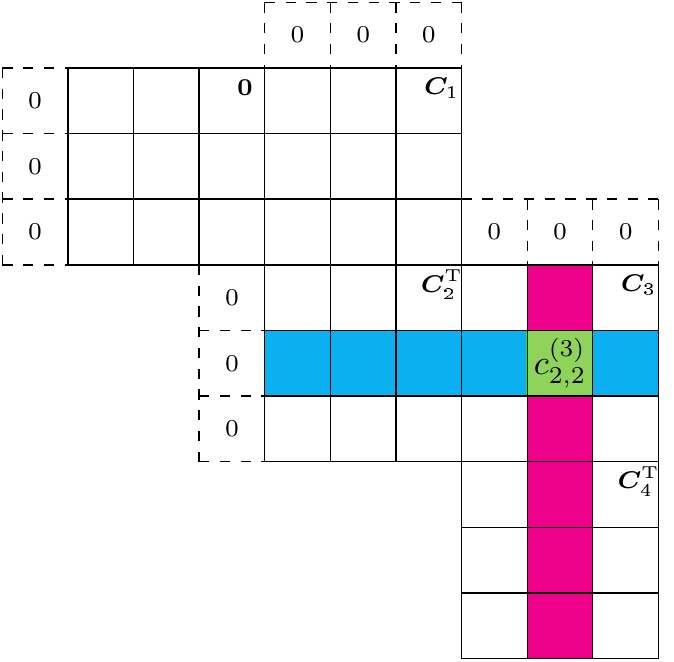}
	\caption{PC and staircase codes with $\nc=7$ and $\nc=6$ and the corresponding code bits $c_{2,2}$ and $c^{(3)}_{2,2}$. The shortened bits are dashed.}
	\label{fig:PCSimpGraph}
	\vspace{-3mm}
\end{figure}

Let $\mathcal{C}$ be an $(\nc, \kc, \dmin)$ binary linear code, where
$\nc$, $\kc$, and $\dmin$ are the code length, dimension, and minimum Hamming distance, respectively. A (two-dimensional) PC with parameters $(\nc^2,\kc^2,\dmin^2)$ and  code rate
$R = \kc^2/\nc^2$  based on component code $\mathcal{C}$ is defined as the
set of all $\nc\times\nc$ arrays such that each row and column of the
array is a codeword of $\mathcal{C}$. A codeword of the PC can thus be represented as a binary matrix $\cc=[c_{i,j}]$ of size $\nc \times \nc $. Fig.~\ref{fig:PCSimpGraph} shows the code array of a PC with component codes of length $\nc=7$, where the code bit $c_{2,2}$ and the row and column constraints in which participates are highlighted. 

\begin{figure*}[t] \centering 
	\includegraphics[align=c,width=4.4cm]{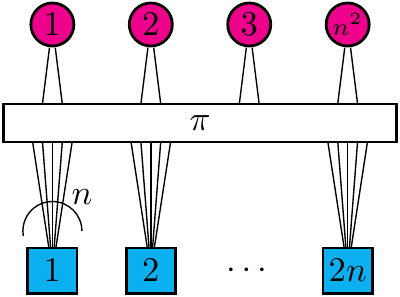}
	\hspace{0.5cm}
    \includegraphics[align=c,width=10.5cm]{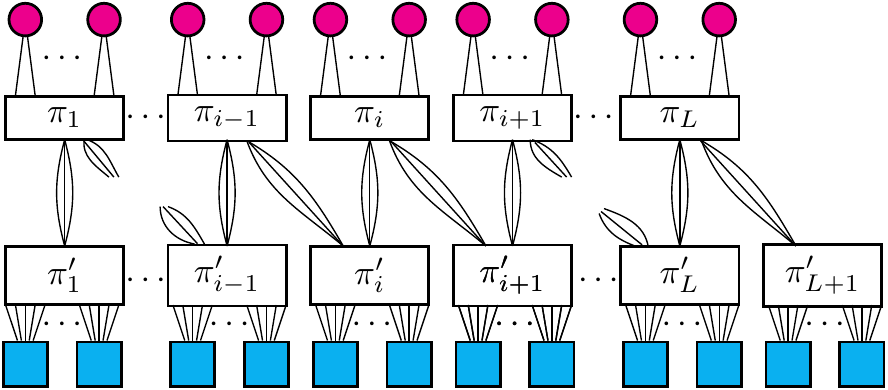} 
	\vspace{1ex}
	\caption{(Left) Schematic of the GLDPC code ensemble containing PCs with component codes of length $n$ as a special case. Circles and squares correspond VNs and CNs, respectively, and $\pi$ denotes the interleaver. (Right) The schematic of the SC-GLDPC code ensemble with $L$ spatial positions and $u=2$ containing staircase codes as a special case. $\pi_i$ and $\pi^{\prime}_i$ represent the random permutations for VNs and CNs, respectively, at spatial position $i$.} 
	\label{ensamble} 
\end{figure*}

PCs can be represented by a Tanner graph, where variable nodes (VNs) represent code bits and constraint nodes (CNs) represent row and column codes, respectively. For a PC with component code length $n$, the corresponding Tanner graph has $n^2$ degree-$2$ VNs (each of the $n^2$ code bits participates in $2$ code constraints, a row constraint and a column constraint) and $2n$ degree-$n$ CNs ($n$ bits participate in each code constraint). PCs are contained in the ensemble of GLDPC codes described by the Tanner graph in Fig.~\ref{ensamble}, where each CN corresponds to a block code of length $n$ (the component code of the PC). Note that the Tanner graph of a PC is a particular instance of the Tanner graph in Fig.~\ref{ensamble}, where the connectivity between VNs and CNs is  deterministic.

A binary staircase code is defined by a two-dimensional code array that has the form of a staircase. Formally, given a component code of length $n$, a staircase code is defined as the set of all matrices $\Bmat_i$ of size ${{\frac{\nc}{2}} \times {\frac{\nc}{2}}}$, $i=1,2,\ldots$, such that each row of the matrix $[\Bmat_{i-1}^\rmT,\Bmat_i]$ is a codeword of $\mathcal{C}$. $\Bmat_0$ is initialized to all zeros and the code rate is $R = 1- \frac{2(\nc-\kc)}{\nc}$ \cite{staircase_frank}. We will refer sometimes to the matrices $\Bmat_i$ as blocks.  In Fig.~\ref{fig:PCSimpGraph},  we plot the code array corresponding to the first four spatial positions of a staircase code with component codes of length $n=6$. The code bit $c^{(3)}_{2,2}$ and the row and column constraints in which participates are  highlighted. 

A staircase code can be seen as a class of \emph{spatially coupled codes} and as such it can be interpreted as spanning over a number of spatial positions, where the code bits in position $i$ correspond to the code bits $[\Bmat_{i-1}^\rmT,\Bmat_i]$, i.e., each spatial position contains two blocks of code bits. We will denote the code bits in position $i$ by the matrix $\Cmat^{(i)}\triangleq [\Bmat_{i-1}^\rmT,\Bmat_i]=[c^{(i)}_{p,j}]$, of dimensions $\frac{n}{2}\times n$. 

As PCs, staircase codes can also be represented by a Tanner graph and as such can be seen as an instance of a SC-GLDPC code ensemble. The Tanner graph of a SC-GLDPC code ensemble of coupling memory $u$ is constructed by placing $L$ copies of a regular GLDPC code of VN degree $\dv$ and CN
degree $\dc$ in $L$ spatial positions in the set $\mathcal{L}=\{1,\ldots,L\}$. Each spatial position consists of $M$ CNs and $N$ VNs. The $L$ copies are then coupled as follows: each VN at position $i\in \mathcal L$ is connected to $\dv$
CNs in positions in the range $[i,\ldots,i+u-1]$, chosen uniformly at random. Likewise, each CN at position $i\in \mathcal L$ is connected to
$\dc$ randomly chosen VNs at positions in the range $[i-u+1,\ldots,i]$. This chain of coupled codes may be terminated by appending $u-1$ spatial positions with CNs only at the end of the chain. The Tanner graph of a staircase code is contained in the Tanner graph of a SC-GLDPC code ensemble with VN degree $2$, coupling memory $u=2$, and $N=\frac{n^2}{2}$ degree-$2$ VNs and
and $M=n$ degree-$n$ CNs per spatial position. The Tanner graph of such a SC-GLDPC code ensemble, originally considered in \cite{JianPfister2017},  is  depicted in Fig.~\ref{ensamble}.


The connection between PCs and GLDPC codes and between staircase codes and SC-GLDPC  codes, allows the use of tools for the analysis of codes-on-graphs, such as density evolution, to analyze PCs.



\subsection{Channel Model}

We assume transmission over the binary-input additive white Gaussian
noise (bi-AWGN) channel. For simplicity, for the definitions below we assume transmission using a PC. The
channel output corresponding to code bit $c_{i,j}$ 
is given by
\begin{align}\label{channel_inst}
y_{i,j}=x_{i,j}+z_{i,j},
\end{align} 
where $x_{i,j}=(-1)^{c_{i,j}}$, $z_{i,j}\sim \mathcal{N}(0,\sigma^2)$, $\sigma^2=(2 R
E_\mathrm{b}/N_0)^{-1}$ and $E_\mathrm{b}/N_0$ is the signal to noise ratio. We denote by $\lalone=[L_{i,j}]$ the matrix of channel
log-likelihood ratios (LLRs) and by $\rr=[r_{i,j}]$ the matrix of hard
decisions at the channel output, where $r_{i,j}$ is obtained by
computing the sign of $L_{i,j}$ and mapping $+ 1 \mapsto 0$ and $- 1
\mapsto 1$. We denote this mapping by $\BB(\cdot)$, i.e.,
$r_{i,j}=\BB(L_{i,j})$. With some abuse of notation, we also write
$\rr=\BB(\lalone)$. 

For the case of staircase codes, the matrices of LLRs and hard decisions must be defined per each spatial position, i.e., we define $\lalone^{(i)}$ and $\rr^{(i)}$, of dimensions $\frac{n}{2}\times n$, in accordance to $\Cmat^{(i)}$.

Remark: The target application for the proposed decoding algorithm is very high throughput applications. A salient application is  next generation fiber-optic links, which will support throughputs up to 1 Tb/s. The optical channel in practical scenarios such as wavelength division multiplexing  can be approximated with the well-established Gaussian noise (GN) model \cite{Poggiolini2012}. The gist of the GN model is that the interplay between Kerr nonlinearity and chromatic dispersion in the optical channel  can be accurately modeled as a Gaussian noise under some mild conditions. Thus, the bi-AWGN channel considered in this paper is also relevant for this particularly interesting application.

\subsection{Bounded Distance Decoding}
\label{BDD}



Consider now the decoding of an arbitrary row or column component
code, assuming that the codeword $\lc=(c_1,\ldots,c_{\nc})$ is
transmitted and decoding is based on the hard-detected bits  at the channel output,
$\lr=(r_1,\ldots,r_{\nc})$. BDD corrects all
error patterns with Hamming weight up to the error-correcting
capability of the code $t=\left\lfloor\frac{\dmin-1}{2}\right\rfloor$. If the
weight of the error pattern is larger than $t$ and there exists
another codeword $\tilde{\lc} \in \mathcal{C}$ with
$\mathsf{d_H}(\tilde{\lc},\lr)\le t$, then BDD erroneously decodes $\lr$ onto
$\tilde{\lc}$ and a so-called \emph{miscorrection} occurs.  Otherwise, if
such codeword does not exist, BDD fails and we use the convention that the
decoder outputs $\lr$. Thus, the decoded vector $\hat{\rbol}$ for BDD
can be written as
\begin{equation}
\label{eq:BDD_VN}
\scalemath{0.912}{	\hat{\rbol} = \begin{cases}
	\lc & \text{if } \ham(\lc, \rbol)  \leq t \\
	\tilde{\lc} \in \mathcal{C} & \text{if } \ham(\lc, \rbol) > t \text{ and
	} \exists{\tilde{\lc}} \; \text{such that} \;\ham(\tilde{\lc}, \rbol) \leq t \\
	\rbol & \text{otherwise}
	\end{cases}.}
\end{equation}

The decoding of product-like codes can then be accomplished in an iterative fashion based on  BDD of the component codes and iterating between the row and column decoders. Here, we refer to iterative decoding of product-like codes based on BDD of the component codes as iBDD.

\section{Iterative Bounded Distance Decoding With Scaled Reliability}
\label{BDD-SRsec}

In this section, we propose a modification of the conventional iBDD of product-like codes. The main idea is to exploit the channel reliabilities in the decoding of the component codes, while only binary messages are exchanged between the component decoders. We refer to this algorithm as iBDD with scaled reliability (iBDD-SR). The proposed algorithm applies to product-like codes in general, including, e.g., PCs, half PCs \cite{Justesen2011}, staircase codes, and braided codes. For illustration purposes, we focus on PCs and staircase codes.

\subsection{Iterative Bounded Distance Decoding with Scaled Reliability for Product Codes}\label{iBDD-SRPC}  

Without loss of generality, assume that decoding starts with the decoding of the row codes and, in particular, consider the decoding of the $i$th row code at iteration $\ell$. 

Let $\bm{\Psi}^{\mathsf{c},(\ell-1)}=[\psi_{i,j}^{\mathsf{c},(\ell-1)}]$ be the matrix of hard decisions on code bits $c_{i,j}$ after the decoding of the $\nc$ column codes at iteration $\ell-1$. Row decoding is then performed based on $\bm{\Psi}^{\mathsf{c},(\ell-1)}$. First, BDD  is performed.  The output of the BDD stage of the $i$th row component code  corresponding to code bit $c_{i,j}$, denoted by $\bar\mu_{i,j}^{\mathsf r}$, takes values on a ternary alphabet, $\bar\mu_{i,j}^{\mathsf r} \in \{\pm1, 0 \}$, where the decoded bits are mapped according to $0\mapsto +1$ and $1\mapsto -1$ if BDD is successful, and the output is $0$ if a decoding fails.  

The reliability information on code bit $c_{i,j}$ is then formed
according to
\begin{equation}\label{eq:GMDchrel_VN_scale}
\mu_{i,j}^{\mathsf r, (\ell)}=\w_i^{\mathsf r, (\ell)} \cdot \bar{\mu}_{i,j}^{\mathsf r,
(\ell)} + L_{i,j}, 
\end{equation}
where $\w_i^{\mathsf r, (\ell)} $ is a scaling factor corresponding to the reliability of the decision on $c_{i,j}$ at the output of the BDD (it may be different for each iteration) that should be properly optimized. In \cite{She18} we optimized the scaling factors for PCs based on simulations. Unfortunately a simulation-based approach becomes infeasible for spatially-coupled product-like codes such as staircase codes. In Section~\ref{sec:DE_PCs}, we show that the scaling factors  can be found via density evolution, which renders the search (and hence the decoder design) feasible.

Finally, the hard decision on code bit $c_{i,j}$ made by the $i$th row decoder is formed as
\begin{equation}\label{eq:BDDchrel_VN}
\psi_{i,j}^{\mathsf{r},(\ell)}=
\BB(\mu_{i,j}^{\mathsf r, (\ell)}) , 
\end{equation}
where ties can be broken with any policy.

\begin{figure}[!t] \centering 
 	\includegraphics[width=\columnwidth]{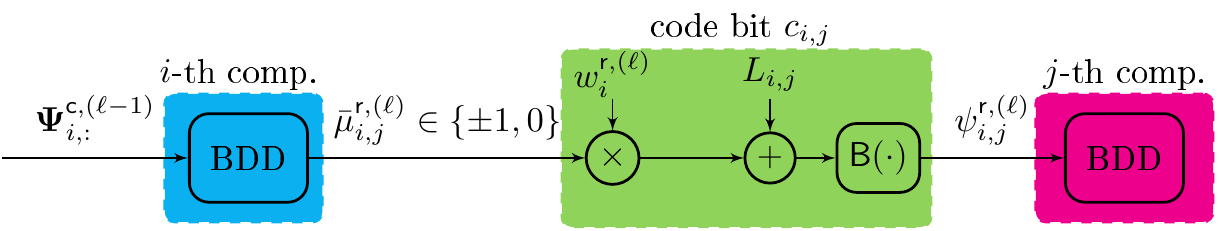}  
 	\caption{Block diagram of iBDD-SR for PCs.}  
 	\label{SysPCCST} 
	\vspace{-2ex}
 \end{figure}

The hard decision $\psi_{i,j}^{\mathsf{r},(\ell)}$ is the message
passed from the $i$-th row code to the $j$-th column code. In
particular, after applying this procedure to all row codes, the matrix
$\boldsymbol{\Psi}^{\mathsf{r},(\ell)}=[\psi_{i,j}^{\mathsf{r},(\ell)}]$
is formed and used as the input for the $n$ column decoders, and column decoding based on
$\boldsymbol{\Psi}^{\mathsf{r},(\ell)}$ is performed. As before, we assume that the output of the BDD stage of the $j$th column component code corresponding to code bit $c_{i,j}$, denoted by $\bar\mu_{i,j}^{\mathsf c}$, takes values on $\{\pm1, 0 \}$.  Then, the hard  decision on $c_{i,j}$ made by the $j$th column decoder at iteration $\ell$ is formed as
\begin{equation*}\label{eq:BDDchrel_VN_scale}
\scriptsize{
		\psi_{i,j}^{\mathsf{c},(\ell)}=\BB(\mu_{i,j}^{\mathsf c, (\ell)})},
\end{equation*}
where $\mu_{i,j}^{\mathsf c, (\ell)}=\w^\col_\ell \cdot \bar\mu_{i,j}^{\mathsf c, (\ell)} + L_{i,j}$ is the reliability of code bit $c_{i,j}$.
After decoding of the $\nc$ column codes at decoding iteration $\ell$, the matrix $\bm{\Psi}^{\mathsf{c},(\ell)}=[\psi_{i,j}^{\mathsf{c},(\ell)}]$ is passed to the $\nc$ row decoders for the next decoding iteration. The iterative process continues until a maximum number of iterations is reached. The iBDD-SR of PCs is schematized in Fig.~\ref{SysPCCST}.

The crucial modification in iBDD-SR with respect to conventional iBDD is that the hard decisions passed between component decoders are not simply the result of the BDD of the component codes, but are made on the sum of a scaled version of the output of the BDD decoder and the channel LLR.  
Therefore, the channel reliabilities are exploited to make the final hard decisions at each row and column decoding stage. 
Intuitively, since the channel reliability is exploited in the hard decisions at each row and column decoding, in the case that the reliability of the channel is high and conventional iBDD introduces miscorrections, the modified algorithm may combat the possible miscorrections. 


\subsection{Iterative Bounded Distance Decoding with Scaled Reliability for Staircase Codes}\label{iBDDSR_ST}
Similar to PCs, staircase codes can be decoded iteratively. Decoding of staircase codes is typically performed in a sliding-window fashion, iterating between the row and column decoders within a window containing a number of staircase blocks and after a given number of iterations shifting the window by one staircase block \cite{staircase_frank}.  Thus, the iBDD-SR of PCs described in the previous subsection readily extends to staircase codes. However, in this case, the scaling factors may be different for each spatial position, and this needs to be considered when optimizing them.

\section{Density Evolution of iBDD-SR}
\label{sec:DE_PCs}

In this section, we provide the density evolution analysis for GLDPC codes and SC GLDPC codes, which  
contain PCs and staircase codes as particular cases for a certain choice of parameters. The analysis provides, as a byproduct, the optimal scaling factors of iBDD-SR (optimal in an asymptotic sense, for infinitely long block length \cite{Lechner2012}). Note that for finite block length, the scaling factors derived from the density evolution analysis are not necessarily the ones that minimize the error probability.  However, selecting the scaling factors based on density evolution avoids resorting to an optimization based on Monte-Carlo simulations, which may be very time-consuming, and is unfeasible for staircase codes. Furthermore, interestingly, comparing the performance of iBDD-SR with scaling factors derived from the density evolution to that with scaling factors found by simulations\footnote{Note that due to the limitation of resources, a full exhaustive search is infeasible, i.e., only a grid search with a certain accuracy (step size) is possible.}, we have observed that the finite length performance of the former is always slightly better.

As discussed in Section~\ref{PCstaircase},
PCs and staircase codes are contained in a particular GLDPC and SC-GLDPC code ensemble, respectively, whose asymptotic behavior can be rigorously analyzed via density evolution. In \cite{JianPfister2017}, the density evolution of iBDD for transmission over the binary symmetric channel (BSC) was derived. In this section, we derive the density evolution of iBDD-SR for GLDPC code ensembles and SC-GLDPC code ensembles by extending the density evolution of conventional iBDD in \cite{JianPfister2017} to the iBDD-SR case. It is worth mentioning that product-like codes are \emph{structured} codes and therefore their decoding threshold is not necessarily fully characterized by the threshold of the corresponding GLDPC code ensemble. In \cite{Haeger2017tit}, the rigorous density evolution analysis that characterizes the asymptotic decoding performance over the binary erasure channel (BEC) of a deterministic construction of product-like codes  that encompasses several classes of product-like codes such as irregular PCs, block-wise braided codes, and staircase codes, was derived. Its extension to the BSC, however, is very cumbersome. In practice, for the BEC the thresholds predicted by the rigorous density evolution and those predicted by the density evolution of the corresponding GLDPC ensemble are virtually identical \cite{Haeger2017tit,Hager15b}. Thus, in this paper we opt to derive the density evolution of iBDD-SR for the GLDPC and SC-GLDPC ensembles encompassing PCs and staircase codes.

\begin{figure}[!t] \centering 
	\includegraphics[width=\columnwidth]{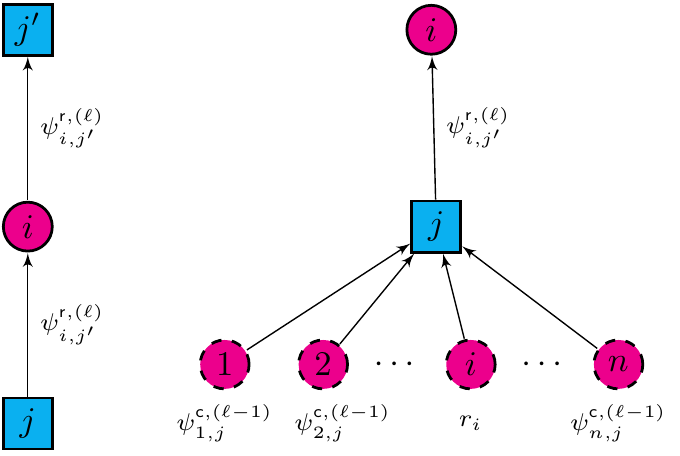}  
	\vspace{-2ex}
	\caption{Block diagram of the message passing algorithm for GLDPC codes. Circles and squares correspond to  VNs and CNs, respectively.}  \vspace{-2ex}
	\label{Messagepass} 
\end{figure}

The GLDPC codes that we consider here can be represented graphically by a Tanner graph where each VN is connected to two CNs, corresponding to one row and one column code (see Section~\ref{PCstaircase} and Fig.~\ref{ensamble}). We consider 
the iterative hard decision decoding algorithm based on extrinsic BDD of the component codes proposed in \cite{JianPfister2017}. In particular, at the $i$-th VN, the input message from the $j$-th CN is forwarded to the $j'$-th CN. At the $j$-th CN, the input corresponding to the $i$-th VN is replaced by the hard-detected bit $r_i$ at the channel output. This message passing algorithm is illustrated in Fig.~\ref{Messagepass}. Note that substituting the $i$-th VN message with $r_i$ makes the decoding rule extrinsic, i.e., the decision on the $i$-th VN does not depend on previous decisions inside the decoder. The extrinsic decoding rule is essential for the density evolution analysis. 

\subsection{Density Evolution Analysis of iBDD-SR for GLDPC Code Ensembles}\label{sec:DE_PCs}

We consider transmission over the bi-AWGN channel and analyze the decoder behavior by assuming the transmission of the all-zero codeword. At the CNs, we assume a length-$n$ binary component code with error-correcting capability $t$.

We denote the channel output error probability, i.e., the error probability that would be attained by applying a hard detection to the bi-AWGN channel output, as $p_\mathsf{ch}$.  
 For an arbitrary row/column BDD stage, we denote by $\Pue\left( i \right)$ the probability that a randomly selected bit in the component code's codeword is decoded incorrectly when it was initially in error and there are $i$  errors in the other $n-1$ positions, and by $\Puc\left( i \right)$  the probability that a randomly selected bit in the component code's codeword is decoded correctly when it was initially in error and there are $i$  errors in the remaining $n-1$ positions. The probability that a randomly selected bit in the component code's codeword is erased when it was initially in error and there are $i$ errors in the remaining $n-1$ positions is denoted by $P^\epsilon(i)$.  We have obviously that $P^\epsilon\left( i \right)=1-{\Pue}\left( i \right)-{P^{\mathsf c}}\left( i \right)$. Similarly, we denote by $\Que\left( i \right)$, $\Quc\left( i \right)$, and $Q^\epsilon(i)$ 
the probability that a randomly selected bit in the component code's codeword is decoded incorrectly, correctly, and erased, respectively, when the bit was initially correct and there are $i$ errors in the remaining $n-1$ positions. Note that  $Q^\epsilon\left( i \right)=1-\Que\left( i \right)-\Quc\left( i \right)$.

We have that
\begin{align}
\label{Pe}
	{\Pue}\left( i \right) = \begin{dcases}
	0 &
	\text{if} \;\;  0 \le i \le t - 1  \\
	\mathop {\sum\limits_{\delta  = 1}^t {\sum\limits_{j = 0}^{\delta} } } \frac{{{h}+1}}{n}{A_{h+1}}F^{(1)}_{h} & 
	\text{if} \;\; t\le i \le n-t-2\\
	1 &
	\text{otherwise} \\
	\end{dcases}
\end{align}
and
\begin{align}
\label{Qc}
	\Quc\left( i \right) = \begin{dcases}
	1 &
	\text{if} \;\;  0 \le i \le t  \\
	\mathop {\sum\limits_{\delta  = 1}^t {\sum\limits_{j = 0}^{\delta} } } \frac{n-{{h}}}{n}{A_{h}}F^{(1)}_{{h}} & 
	\text{if} \;\; t+1\le i \le n-t-1\\
	0 &
	\text{otherwise}.  \\
	\end{dcases}
\end{align}
where $h=i-\delta+2j$ and $F^{(1)}_{h}$ is defined as
\begin{align}\label{f_pe} 
F^{(1)}_{h} \triangleq \frac{{h \choose {h-j}}{{n-h-1} \choose {\delta-j}}}{{{n-1} \choose i}}.
\end{align} 
In \eqref{Pe}, $A_h$ is the weight enumerator of the component code, which can be tightly approximated as
\begin{align}\label{weight_dist} 
	A_h\, \approx\, \begin{dcases}
	2^{-\nu t} {n \choose h} \Big( 1+o(1) \Big) & 
	\text{if} \;\; 2t+1\le h \le n-2t-1\\
	1 & 
	\text{if} \;\; h=0, h=n\\
	0 &
	\text{otherwise}.  \\
	\end{dcases}
\end{align}
Similarly, one can compute ${P^{\mathsf c}}\left( i \right)$ and  $\Que\left( i \right)$ as
\begin{align}
\label{Pc}
	{\Puc}\left( i \right) = \begin{dcases}
	1 &
	\text{if} \;\;  0 \le i \le t - 1  \\
	\mathop {\sum\limits_{\delta  = 1}^t {\sum\limits_{j = 0}^{\delta-1} } } \frac{n-{h}}{n}{A_{h}}F^{(2)}_{h} & 
	\text{if} \;\; t\le i \le n-t-2\\
	0 &
	\text{otherwise}  \\
	\end{dcases}
\end{align}
and
\begin{align}
\label{Qe}
	\Que\left( i \right) = \begin{dcases}
	0 &
	\text{if} \;\;  0 \le i \le t   \\
	\mathop {\sum\limits_{\delta  = 1}^t {\sum\limits_{j = 0}^{\delta-1} } } \frac{{{h}+1}}{n}{A_{{h}+1}}F^{(2)}_{{h}} & 
	\text{if} \;\; t+1\le i \le n-t-1\\
	1 &
	\text{otherwise}  \\
	\end{dcases}
\end{align}
where ${h}=i-\delta+2j+1$ and $F^{(2)}_{h}$ is defined as
\begin{align}\label{f_pc} 
F^{(2)}_{h} \triangleq \frac{{h \choose {h-j}} {{n-h-1} \choose {\delta-j-1}}}{{{n-1}\choose i}}.
\end{align}

For ease of understanding, we briefly explain the derivation of ${\Pue}\left( i \right)$ in \eqref{Pe} in the appendix. The derivation of $\Quc(i)$, $\Puc(i)$, and $\Que(i)$ follows a similar reasoning. 

For the considered GLDPC code ensemble, CNs are divided into two sets of equal size to capture the serial row/column decoding schedule of PCs. Each set defines a CN type. We refer to the two CN types as row and column CN types. Each VN is connected to one row-type CN and to one column-type CN. Each decoding iteration consists of a row CN elaboration, followed by a column CN elaboration.
In the following, we denote by $\mep$ the error probability associated to the messages exchanged by the component decoders. In particular, we denote by $\mep^{\row,(\ell)}$ and $\mep^{\col,(\ell)}$ the message error probability at the output of the row component decoder (row-type CN) and column component decoder (column-type CN), respectively, at the $\ell$th iteration. The message error probability at the input of a row-type CN at the $\ell$th iteration is given by $\mep^{\col,(\ell-1)}$, whereas the message error probability at the input of a column-type CN during the $\ell$th iteration is $\mep^{\row,(\ell)}$.
At the first iteration, we have $\mep^{\col,(0)}=p_\mathsf{ch}$, i.e., the input of the row-type CNs is initialized with the channel observations.

In iBDD-SR, the output of the BDD decoder is from a ternary alphabet, i.e., ${\bar\mu}_{i,j}^{\mathsf r, (\ell)} \in  \{\pm1, 0 \}$ (see Section~\eqref{iBDD-SRPC}), where (under the all-zero codeword assumption) ${\bar\mu}_{i,j}^{\mathsf r, (\ell)}=0$, ${\bar\mu}_{i,j}^{\mathsf r, (\ell)}=-1$, and ${\bar\mu}_{i,j}^{\mathsf r, (\ell)}=+1$ correspond to  erasure (failure),  erroneous decoding, and  correct decoding, respectively. In the following, the probabilities of ${\bar\mu}_{i,j}^{\mathsf r, (\ell)}=0$, ${\bar\mu}_{i,j}^{\mathsf r, (\ell)}=-1$, and ${\bar\mu}_{i,j}^{\mathsf r, (\ell)}=+1$ conditioned on $x_{i,j}=+1$ are denoted as $f^{\epsilon}(\mep;p_\mathsf{ch})$, $\fue(\mep;p_\mathsf{ch})$, and $\fuc(\mep;p_\mathsf{ch})$, respectively, for a given input message error probability $\mep$ and a channel error probability $p_\mathsf{ch}$. One can check that
\begin{align}\label{fe} 
\nonumber \fue(\mep;p_\mathsf{ch})  = \sum\limits_{i = 0}^{n - 1} &  {\Big( {\begin{array}{*{20}{c}}
		{n - 1}\\
		i
		\end{array}} \Big)} {\mep^i}{\left( {1 - \mep} \right)^{n - i - 1}} \cdot \\  & \Big( {p_\mathsf{ch}{\Pue}\left( i \right) + {{\bar p}_\mathsf{ch}}\Que\left( i \right)} \Big),
\end{align}
\begin{align}\label{fc} 
\nonumber \fuc(\mep;p_\mathsf{ch})  = \sum\limits_{i = 0}^{n - 1} &  {\Big( {\begin{array}{*{20}{c}}
		{n - 1}\\
		i
		\end{array}} \Big)} {\mep^i}{\left( {1 - \mep} \right)^{n - i - 1}} \cdot \\  & \Big( {p_\mathsf{ch}{P^{\mathsf c}}\left( i \right) + {{\bar p}_\mathsf{ch}}\Quc\left( i \right)} \Big),
\end{align}
\begin{align}\label{fepsilon} 
\nonumber f^{\epsilon}(\mep;p_\mathsf{ch})  = \sum\limits_{i = 0}^{n - 1} &  {\Big( {\begin{array}{*{20}{c}}
		{n - 1}\\
		i
		\end{array}} \Big)} {\mep^i}{\left( {1 - \mep} \right)^{n - i - 1}} \cdot \\  & \Big( {p_\mathsf{ch}{P^{\epsilon}}\left( i \right) + {{\bar p}_\mathsf{ch}}{Q^{\epsilon}}\left( i \right)} \Big),
\end{align}
where ${\bar p}_\mathsf{ch}\triangleq 1-{p}_\mathsf{ch}$. 

Under the all-zero codeword assumption, one can readily find that $L_{i,j}\sim\mathcal{N}(2/\sigma^2,4/\sigma^2)$. Note that for the density evolution, as all row CNs and all column CNs are of the same type (i.e., they behave identically), we can assume that $\w^{\row,(\ell)}_i=\w^{\row,(\ell)}$ and  $\w^{\col,(\ell)}_i=\w^{\col,(\ell)}$ for all $i$. We are now interested in finding $\mep^{\row,(\ell)}\triangleq p( \w^{\row,(\ell)} \cdot {\bar\mu}_{i,j}^{\mathsf r, (\ell)} + L_{i,j} < 0 )$ (alternatively $\mep^{\col,(\ell)} \triangleq p( \w^{\col,(\ell)} \cdot {\bar\mu}_{i,j}^{\mathsf c, (\ell)} + L_{i,j} < 0 )$), as the iBDD-SR output error probability for the row (column) decoding. To proceed further, one needs to find the distribution of $\w^{\row,(\ell)} \cdot {\bar\mu}_{i,j}^{\mathsf r, (\ell)} + L_{i,j}$. The task of finding this distribution is rendered complicated by observing that the messages ${\bar\mu}_{i,j}^{\mathsf r, (\ell)}$ and  ${\bar\mu}_{i,j}^{\mathsf c, (\ell)}$ are statistically dependent on $L_{i,j}$. In fact, the evolution of the error probability at the output of the CNs discussed so far is based on the approach proposed in \cite{JianPfister2017}, where the CN operation is modified (with respect to the classical behavior) to account for the channel observation $r_{i,j}$ when computing the extrinsic terms ${\bar\mu}_{i,j}^{\mathsf r, (\ell)}$ and ${\bar\mu}_{i,j}^{\mathsf c, (\ell)}$. In our model, the BDD output message is modified according to the sum $\w^{\row,(\ell)} \cdot {\bar\mu}_{i,j}^{\mathsf r, (\ell)} + L_{i,j}$, hence the channel observation for code bit $c_{i,j}$ is used at both the BDD input and output. Here instead of finding directly the distribution of $\w^{\row,(\ell)} \cdot {\bar\mu}_{i,j}^{\mathsf r, (\ell)} + L_{i,j}$, we first expand $\mep^{\row,(\ell)}$ using the auxiliary RV $\hat L_{i,j}$ giving the sign of $L_{i,j}$, i.e., $\hat L_{i,j} = \text{sign}(L_{i,j})$. Employing Bayes' rule,  $\mep^{\row,(\ell)}$ can be written as
\begin{align}\label{p_bays} 
\mep^{\row,(\ell)} &= p(\mu_{i,j}^{\mathsf r, (\ell)}<0) \nonumber \\ &= \sum\limits_{\mathclap{\substack{
		{\bar \mu _{i,j}^{r,(\ell )}\in \{ 0, \pm 1\} }\\
			{{{\hat L}_{i,j}}\in \{  \pm 1\} }}}} {p(\mu_{i,j}^{\mathsf r, (\ell)} < 0|\bar \mu _{i,j}^{r,(\ell )},{{\hat L}_{i,j}})} p(\bar \mu _{i,j}^{r,(\ell )}|{\hat L_{i,j}})p({\hat L_{i,j}}).
\end{align}
It is easy to see that  $L_{i,j} \to \hat L_{i,j} \to \bar\mu _{i,j}^{r,(\ell )}$ form a Markov chain, hence, conditioned on $\hat L_{i,j}$, $\bar\mu_{i,j}^{r,(\ell )}$ is independent of $L_{i,j}$. Using this,
\begin{align*}
p(\mu_{i,j}^{\mathsf r, (\ell)} < 0|&\bar \mu _{i,j}^{r,(\ell )},{{\hat L}_{i,j}})\\&=p(\w_i^{\mathsf r, (\ell)} \cdot \bar{\mu}_{i,j}^{\mathsf r,
(\ell)} + L_{i,j}<0|\bar \mu _{i,j}^{r,(\ell )},{{\hat L}_{i,j}})\\
&=p(L_{i,j} < - \w_i^{\mathsf r, (\ell)} \cdot \bar{\mu}_{i,j}^{\mathsf r,
(\ell)} |{\hat L}_{i,j})
\end{align*}
One can easily check that $p(L_{i,j} < - \w^{\row,(\ell)} |{\hat L}_{i,j}=1)=p(L_{i,j} <0 |{\hat L}_{i,j}=1)=0$ and $p(L_{i,j} < - \w^{\row,(\ell)} |{\hat L}_{i,j}=-1)=p(L_{i,j} <0 |{\hat L}_{i,j}=-1)=1$. Using this in \eqref{p_bays} yields 
\begin{align}\label{p_simplify} 
 \mep^{\row,(\ell)} = & p(0<L_{i,j} < \w^{\row,(\ell)})p(\bar \mu _{i,j}^{r,(\ell )}=-1|{\hat L_{i,j}}=1)+\nonumber \\ &
p(L_{i,j} < -\w^{\row,(\ell)})p(\bar \mu _{i,j}^{r,(\ell )}=1|{\hat L_{i,j}}=-1)+\nonumber \\ & (1-p(\bar \mu _{i,j}^{r,(\ell )}=1|{\hat L_{i,j}}=-1))p(L_{i,j}=-1).
\end{align}
Further, $p(\bar \mu _{i,j}^{r,(\ell )}=-1|{\hat L_{i,j}}=1)$ and $p(\bar \mu _{i,j}^{r,(\ell )}=-1|{\hat L_{i,j}}=-1)$ can be obtained based on $\Que\left( i \right)$ and $P^{\mathsf c}\left( i \right)$ as 
\begin{align}\label{qe} 
  f^{Q^{\mathsf e}}(\mep) &\triangleq    p(\bar \mu _{i,j}^{r,(\ell )}=-1|{\hat L_{i,j}}=1)  \nonumber \\ & =  \sum\limits_{i = 0}^{n - 1}   {\Big( {\begin{array}{*{20}{c}}
		{n - 1}\\
		i
		\end{array}} \Big)} {\mep^i}{\left( {1 - \mep} \right)^{n - i - 1}} \cdot  {\Que\left( i \right)},
\end{align}
\begin{align}\label{pc} 
 f^{\Puc}(\mep) &\triangleq    p(\bar \mu _{i,j}^{r,(\ell )}=1|{\hat L_{i,j}}=-1)  \nonumber \\ & =  \sum\limits_{i = 0}^{n - 1}   {\Big( {\begin{array}{*{20}{c}}
		{n - 1}\\
		i
		\end{array}} \Big)} {\mep^i}{\left( {1 - \mep} \right)^{n - i - 1}} \cdot  {\Puc\left( i \right)}.
\end{align}
Finally, by substituting \eqref{qe} and \eqref{pc} in \eqref{p_simplify} and using the Gaussian distribution of $L_{i,j}$ to compute $p(0<L_{i,j} < \w^{\row,(\ell)})$ and $p(L_{i,j} < -\w^{\row,(\ell)})$, we can track the evolution of the message error probabilities as
\begin{align}\label{p_eror_r} 
&\mep^{\row,(\ell)}  = f^{Q^{\mathsf e}}(\mep^{\col,(\ell-1)})\cdot\left( {\Q\left( {\frac{1}{\sigma } - \frac{{\sigma {w^{\row,(\ell )}}}}{2}} \right) - p_\mathsf{ch}} \right) \nonumber \\ & +f^{\Puc}(\mep^{\col,(\ell-1)})\cdot \Q\left( {\frac{1}{\sigma } + \frac{{\sigma {w^{\row,(\ell )}}}}{2}} \right) + \left( {1 - {f^{\Puc}(\mep^{\col,(\ell-1)}) }} \right)p_\mathsf{ch}
\end{align} 
 and
\begin{align}\label{p_eror_c} 
& \mep^{\col,(\ell)}  =f^{Q^{\mathsf e}}(\mep^{\row,(\ell)})\cdot \left( {\Q\left( {\frac{1}{\sigma } - \frac{{\sigma {w^{\col,(\ell )}}}}{2}} \right) - p_\mathsf{ch}} \right) \nonumber \\ & + f^{\Puc}(\mep^{\row,(\ell)})\cdot\Q\left( {\frac{1}{\sigma } + \frac{{\sigma {w^{\col,(\ell )}}}}{2}} \right)+ \left( {1 - {f^{\Puc}(\mep^{\row,(\ell)})}} \right)p_\mathsf{ch}.
\end{align}
where $\Q(\cdot)\triangleq\frac{1}{\sqrt{2\pi}}\int_x^\infty \mathrm{e}^{\frac{-\xi^2}{2}}\mathrm{d}\xi$ is the familiar tail distribution function of the standard Gaussian distribution and $p_\mathsf{ch}=\Q\left(\frac{1}{\sigma}\right)$.

One can numerically search for the optimal scaling factors ${w^{\col,(\ell )}}$ and ${w^{\col,(\ell )}}$ in the sense of minimizing $\mep^{\row,(\ell)}$ and $\mep^{\col,(\ell)}$, respectively. Alternatively, by neglecting the statistical dependence between $\bar \mu _{i,j}^{r,(\ell )}$ and $L_{i,j}$ one can approximate $w^{\row,(\ell )}$ and $w^{\col,(\ell )}$ as the LLR of the output of a binary error and erasure channel with error probability  
$\fue$ and erasure probability $f^{\epsilon}$, given as
\begin{align}\label{wrow}
w^{\row,(\ell)}=\log\left(\frac{\fuc(\mep^{\col,(\ell-1)};p_\mathsf{ch})}{\fue(\mep^{\col,(\ell-1)};p_\mathsf{ch})}\right)
\end{align}
and
\begin{align}\label{wcol}
w^{\col,(\ell)}=\log\left(\frac{\fuc(\mep^{\row,(\ell)};p_\mathsf{ch})}{\fue(\mep^{\row,(\ell)};p_\mathsf{ch})}\right).
\end{align}
Employing \eqref{wrow} and \eqref{wcol} yields very similar scaling factors as the ones  obtained performing a numerical search. Furthermore, the code threshold given by the scaling factors in \eqref{wrow} and \eqref{wcol} is roughly the same as the one computed based on the numerically optimized scaling factors. Therefore, \eqref{wrow} and \eqref{wcol} provide a good approximation for $w^{\row,(\ell)}$ and $w^{\col,(\ell)}$, respectively. The obtained scaling factors can then be used in \eqref{eq:GMDchrel_VN_scale} to implement iBDD-SR for a particular finite-length PC.

\subsection{Density Evolution Analysis of iBDD-SR for SC-GLDPC Code Ensembles}\label{DEiBDDSR}

The density evolution for GLDPC codes derived in the previous subsection can be readily extended to SC-GLDPC codes with smoothing parameter $u$. For SC-GLDPC codes, we need to track the probabilities of the messages exchanged in the iterative decoding for each spatial position.  Let $\mep^{(\ell)}_a$ be the average bit error probability from VNs at spatial position $a$ to the connected CNs at spatial positions $[a,a+u-1]$. Also, let $\mep^{{\mathrm{c}},(\ell )}_{a}$ be the average bit error probability from CNs at spatial position $a$ to connected VNs at positions $[a-u+1,a]$. $\mep^{(\ell)}_a$ and $\mep^{{\mathrm{c}},(\ell )}_{a}$ can be calculated as
\begin{align}\label{xl} 
\mep^{{\mathrm{c}},(\ell )}_{a} = \frac{1}{{{u}}}\sum\limits_{g = 0}^{{u} - 1} {\mep_{a - g}^{(\ell )}}, 
\end{align}
\begin{align}\label{xcl} 
\mep_a^{(\ell  + 1)} =  \frac{1}{{{u}}}\sum\limits_{g = 0}^{{u} - 1} &   \left( {\Q\left( {\frac{1}{\sigma } - \frac{{\sigma {w_{a+g}^{(\ell)}}}}{2}} \right) - p_\mathsf{ch}} \right)f^{Q^{\mathsf e}}({\mep^{{\mathrm{c}},(\ell )}_{a+g}}) + \nonumber \\  &   \Q\left( {\frac{1}{\sigma } + \frac{{\sigma {w_{a+g}^{(\ell)}}}}{2}} \right)f^{\Puc}({\mep^{{\mathrm{c}},(\ell )}_{a+g}}) \nonumber + \\  & \left( {1 - {f^{\Puc}({\mep^{{\mathrm{c}},(\ell )}_{a+g}}) }} \right)p_\mathsf{ch}.  
\end{align}
We consider $u=2$ to account for the ensemble containing staircase codes. Thus, combining \eqref{xl} and \eqref{xcl}, we obtain
\begin{align}\label{update_eq}
\mep_{a}^{(\ell  + 1)}=\frac{\mep_{a,0}^{(\ell  + 1)} + \mep_{a,1}^{(\ell  + 1)}}{2},	
\end{align}
where $\mep_{a,0}^{(\ell  + 1)}$ and $\mep_{a,1}^{(\ell  + 1)}$ are given as
\begin{align}\label{update_eq0}
\mep_{a,0}^{(\ell  + 1)} = &  \left( {\Q\left( {\frac{1}{\sigma } - \frac{{\sigma {w_{a}^{(\ell)}}}}{2}} \right) - p_\mathsf{ch}} \right)f^{Q^{\mathsf e}}\left({\frac{\mep_{a}^{(\ell )}+\mep_{a-1}^{(\ell )}}{2}}\right) + \nonumber \\  &  \Q\left( {\frac{1}{\sigma } + \frac{{\sigma {w_{a}^{(\ell)}}}}{2}} \right)f^{\Puc}\left({\frac{\mep_{a}^{(\ell )}+\mep_{a-1}^{(\ell )}}{2}}\right) + \nonumber \\  & \left( {1 - f^{\Puc}\left({\frac{\mep_{a}^{(\ell )}+\mep_{a-1}^{(\ell )}}{2}}\right)} \right)p_\mathsf{ch}, 	
\end{align}
\begin{align}\label{update_eq1}
\mep_{a,1}^{(\ell  + 1)} = &  \left( {\Q\left( {\frac{1}{\sigma } - \frac{{\sigma {w_{a+1}^{(\ell)}}}}{2}} \right) - p_\mathsf{ch}} \right)f^{Q^{\mathsf e}}\left({\frac{\mep_{a}^{(\ell )}+\mep_{a+1}^{(\ell )}}{2}}\right) + \nonumber \\  &  \Q\left( {\frac{1}{\sigma } + \frac{{\sigma {w_{a+1}^{(\ell)}}}}{2}} \right)f^{\Puc}\left({\frac{\mep_{a}^{(\ell )}+\mep_{a+1}^{(\ell )}}{2}}\right) + \nonumber \\  & \left( {1 - f^{\Puc}\left({\frac{\mep_{a}^{(\ell )}+\mep_{a+1}^{(\ell )}}{2}}\right)} \right)p_\mathsf{ch}. 	
\end{align}

Similar to the GLDPC code ensemble, $w_a^{(\ell)}$ and $w_{a+1}^{(\ell)}$ can be obtained as
\begin{align}\label{wcom0}
w_{a}^{(\ell)}=\text{log}\left( \frac{\fuc_{n}\left( {\frac{\mep_{a}^{(\ell )}+\mep_{a-1}^{(\ell )}}{2}};p_\mathsf{ch}\right)}{\fue_{n}\left( {\frac{\mep_{a}^{(\ell )}+\mep_{a-1}^{(\ell )}}{2}};p_\mathsf{ch}\right)}\right),
\end{align}
\begin{align}\label{wcom1}
w_{a+1}^{(\ell)}=\text{log}\left( \frac{\fuc_{n}\left( {\frac{\mep_{a}^{(\ell )}+\mep_{a+1}^{(\ell )}}{2}};p_\mathsf{ch}\right)}{\fue_{n}\left( {\frac{\mep_{a}^{(\ell )}+\mep_{a+1}^{(\ell )}}{2}};p_\mathsf{ch}\right)}\right).
\end{align}

We consider the decoding of staircase codes based on the sliding-window operation. To account for the effect of window decoding, we assume that the width of the window  is $U$, i.e., the window contains $U$ spatial positions. Let $\mathcal{W}_U$ be the set containing the indices of the spatial positions in the current window. In window decoding, the decoder freezes the messages coming from the VNs and CNs outside the window, i.e., the VNs and CNs inside the window are updated based on the information exchanged inside the window and no information comes from the positions outside it. In the particular case of staircase codes, which are contained in the ensemble of SC-GLDPC codes with $u=2$, the first and last positions inside the window do not get any information from the positions outside it. Therefore, one can define $\tilde\mep^{(\ell )}_a$ as
\begin{align}
\label{al_prim_window}
\tilde\mep^{(\ell )}_a =
\begin{dcases}
0 &
\text{if } a \notin {{\cal W}_U}\\
\mep^{(\ell)}_a &
\text{if } a \in {{\cal W}_U} \\
\end{dcases},
\end{align} 
and use it in \eqref{update_eq}--\eqref{wcom1} to find the average bit error probability for the spatial positions within the window.

\section{Complexity Considerations}\label{sec:comp}
A thorough complexity analysis of the iBDD-SR algorithm requires delving into the hardware implementation in order to address aspects such as data bus requirements, impact on the degree of parallelism, etc., that go beyond the scope of this paper. We refer the interested reader to \cite{FougstedtiBDDSR2018}, where an efficient architecture for iBDD-SR of PCs is presented.  
In the following, we provide a high-level discussion of the decoding complexity. In particular, we estimate the additional resources required in terms of memory with respect to iBDD for both PCs and staircase codes. We also compare the complexity of iBDD-SR with that of AD \cite{Hag18} in terms of memory requirements.

Consider a PC with BCH component code $(n,k,\dmin)$ of error correcting capability $t$ decoded over $\ell_\text{max}$ iterations. iBDD-SR requires some extra memory compared to iBDD. With reference to \eqref{eq:GMDchrel_VN_scale}, the decision on code bit $c_{i,j}$ at iteration $\ell$ can be divided into two cases: i) $\psi_{i,j}^{\mathsf{r},(\ell)}=\BB(L_{i,j})$  if $\w_i^{\mathsf r, (\ell)}<|L_{i,j}|$ or BDD fails (i.e., $\bar{\mu}_{i,j}^{\mathsf r, (\ell)}=0$); and ii) $\psi_{i,j}^{\mathsf{r},(\ell)}=\BB(\bar{\mu}_{i,j}^{\mathsf r, (\ell)})$ if BDD is successful and $\w_i^{\mathsf r, (\ell)}>|L_{i,j}|$. By storing the channel LLRs ($L_{i,j}$) and $\w_i^{\mathsf r, (\ell)}$, one can implement \eqref{eq:GMDchrel_VN_scale} with a simple logic comparison. Using $n_{\mathsf q}$ bits for quantization of $L_{i,j}$ and $\w_i^{\mathsf r, (\ell)}$, the additional memory required for iBDD-SR compared to iBDD is $(\ell_\text{max}+n^2)n_{\mathsf q} \approx n^2n_{\mathsf q}$ bits.\footnote{Note that $\ell_\text{max} \ll n^2 $.}

Consider now a staircase code with an (even length) BCH component code decoded using a sliding window of size $U$. We remark that the corresponding SC-GLDPC code ensemble comprises $U-1$ spatial positions, i.e., the window size for the corresponding SC-GLDPC ensemble is $U-1$. In general, the scaling factors may be different for each spatial position (and vary also with the number of iterations). This could be impractical. as many scaling factors would be required to be stored. Interestingly, as discussed in Section~\ref{num_res}, the density evolution for SC-GLDPC code ensembles shows that the scaling factors converge to certain values after few ($U-2$) window slides, after which the scaling factors are identical for each window. Thus few scaling factors need to be stored.
Therefore, the additional memory required for the decoding of one staircase block in iBDD-SR compared to iBDD is $((U-2)(U-1)\ell_\text{max}+\frac{n^2}{4})n_{\mathsf q} \approx \frac{n^2n_{\mathsf q}}{4}$\footnote{Note that the typical window size is in the range of $4$--$7$ blocks, hence $(U-2)(U-1)\ell_\text{max} \ll \frac{n^2}{4}$.}.

We remark that the required memory for both PCs and staircase codes is static, i.e., no switching activities are involved, as the LLRs and scaling factors are not updated in the iterative decoding. This means that the additional memory has a limited cost in terms of energy consumption for the decoder \cite{Fougstedt2018,FougstedtiBDDSR2018}.

The direct complexity comparison with AD algorithm is a non trivial task, as AD can be implemented in various ways (see \cite[Remark~7]{Hag18}). In a high level view, AD assigns a flag to each component code, which indicates the corresponding status in decoding. Furthermore, AD reduces the decoding conflicts between component decoders by keeping track of the conflict locations and preventing the bit flips on the most trusted component codes, called anchors. As the anchors can also be miscorrected, AD allows to backtrack the decisions on anchors. Overall, a memory in the order of $4n$ bits is required for storing the status of the component codes. Also $4nt\left( {\left\lceil {{{\log }_2}n} \right\rceil  + 1} \right)$ bits are required for storing the location of conflicts and backtracking procedure (see \cite[Sec.~VI.~B]{Hag18} for more details).
Unlike the static memory of iBDD-SR, the memory required for AD is dynamic, as the status of component codewords and the location of errors changes during the iterative decoding. The dynamic memory is usually significantly more costly in hardware implementation compared to the static counterpart  \cite{Fougstedt2018}.

Note that in iBDD-SR only binary messages are exchanged between component decoders, hence the internal decoder data flow is the same as that of iBDD. Thus,  iBDD-SR is particularly interesting for very high-throughput applications, for which  the internal decoder data flow is a limiting factor.



%
 \newcommand{\tablehighlight}{}
\begin{table*}[t]	
	\caption{Comparison of different decoding algorithms for PCs and staircase codes with ($255$,$231$,$3$) and ($254$,$230$,$3$) BCH component codes, respectively. The code rate of the PC and the staircase code is $0.820$ and $0.811$, respectively. The $E_\mathrm{b}/N_0$, coding
		gains, and corresponding capacity gaps are measured at $\text{BER} = 10^{-6}$ from the simulations. The values for staircase codes are provided within parenthesis.}
	\centering
	\renewcommand{\arraystretch}{1.2}
	\scalebox{0.96}{	
	\begin{tabular}{ccccccc}
		\toprule
		\tablehighlight{decoding algorithm} &
		\makecell{\tablehighlight{channel}\\ \tablehighlight{reliabilities}} &
		\makecell{\tablehighlight{exchanged} \\
			\tablehighlight{messages}} &
		\makecell{\tablehighlight{$E_\mathrm{b}/N_0$} \\
		\tablehighlight{[dB]}} & \makecell{gain over\\ iBDD [dB]}&
		\makecell{capacity [dB]}&
		\makecell{gap from \\ capacity [dB]}\\
		\midrule
		 iBDD & no & hard & $4.62$ ($4.52$) & - & $3.54$ ($3.46$) (HD) & $1.08$ ($1.06$)  \\
		 AD \cite{Hag18} & no & hard & $4.43$ ($4.25$) & $0.21$ $(0.27)$ & $3.54$ ($3.46$) (HD) & $0.89$ $(0.79)$  \\
		 iBDD-SR & yes & hard  & $4.34$ ($4.21$) & $0.29$ ($0.31$) & $2.23$ ($2.14$) (SD) & $2.11$ ($2.07$)  \\
		 ideal iBDD  & no & hard & $4.31$ ($4.19$) & $0.33$ ($0.33$) & $3.54$ ($3.46$) (HD) & $0.77$ ($0.73$) \\
		\bottomrule
	\end{tabular}
}
	\label{Tabcomp}
\end{table*} 

\begin{table*}[t]
	\caption{Comparison of different decoding algorithms for PCs and staircase codes with ($511$,$484$,$3$) and ($510$,$483$,$3$) BCH component codes, respectively. The code rate of the PC and the staircase code is $0.897$ and $0.894$, respectively. The $E_\mathrm{b}/N_0$, coding
			gains, and corresponding capacity gaps are measured at $\text{BER} = 10^{-6}$ from the simulations. The values for staircase codes are provided within parenthesis.}
	\centering
	\renewcommand{\arraystretch}{1.2}
	\scalebox{0.96}{	
	\begin{tabular}{ccccccc}
		\toprule
	\tablehighlight{decoding algorithm} &
		\makecell{\tablehighlight{channel}\\ \tablehighlight{reliabilities}} &
		\makecell{\tablehighlight{exchanged} \\
			\tablehighlight{messages}} & \makecell{\tablehighlight{$E_\mathrm{b}/N_0$} \\
			\tablehighlight{[dB]}} & \makecell{gain over\\ iBDD [dB]}&
		\makecell{capacity [dB]} &
		\makecell{gap from\\ capacity [dB]}\\
		\midrule
		iBDD & no & hard & $5.18$ ($5.06$) & - & $4.36$ ($4.32$) (HD) & $0.82$ ($0.74$) \\
		AD  \cite{Hag18} & no & hard & $5.01$ ($4.86$) & $0.17$ $(0.21)$ & $4.36$ ($4.32$) (HD) & $0.65$ $(0.54)$ \\
		iBDD-SR  & yes & hard  & $4.93$ ($4.80$) & $0.24$ ($0.265$) & $3.15$ ($3.11$) (SD) & $1.78$ ($1.69$) \\
		ideal iBDD  & no & hard & $4.92$ ($4.79$) & $0.26$ ($0.27$) & $4.36$ ($4.32$) (HD) & $0.56$ ($0.47$) \\
		\bottomrule
	\end{tabular}
 }
	\label{Tabcomp1}
\end{table*}

\section{Numerical  Results}
\label{num_res}

\begin{figure}[!t] \centering 
	\includegraphics[width=\columnwidth]{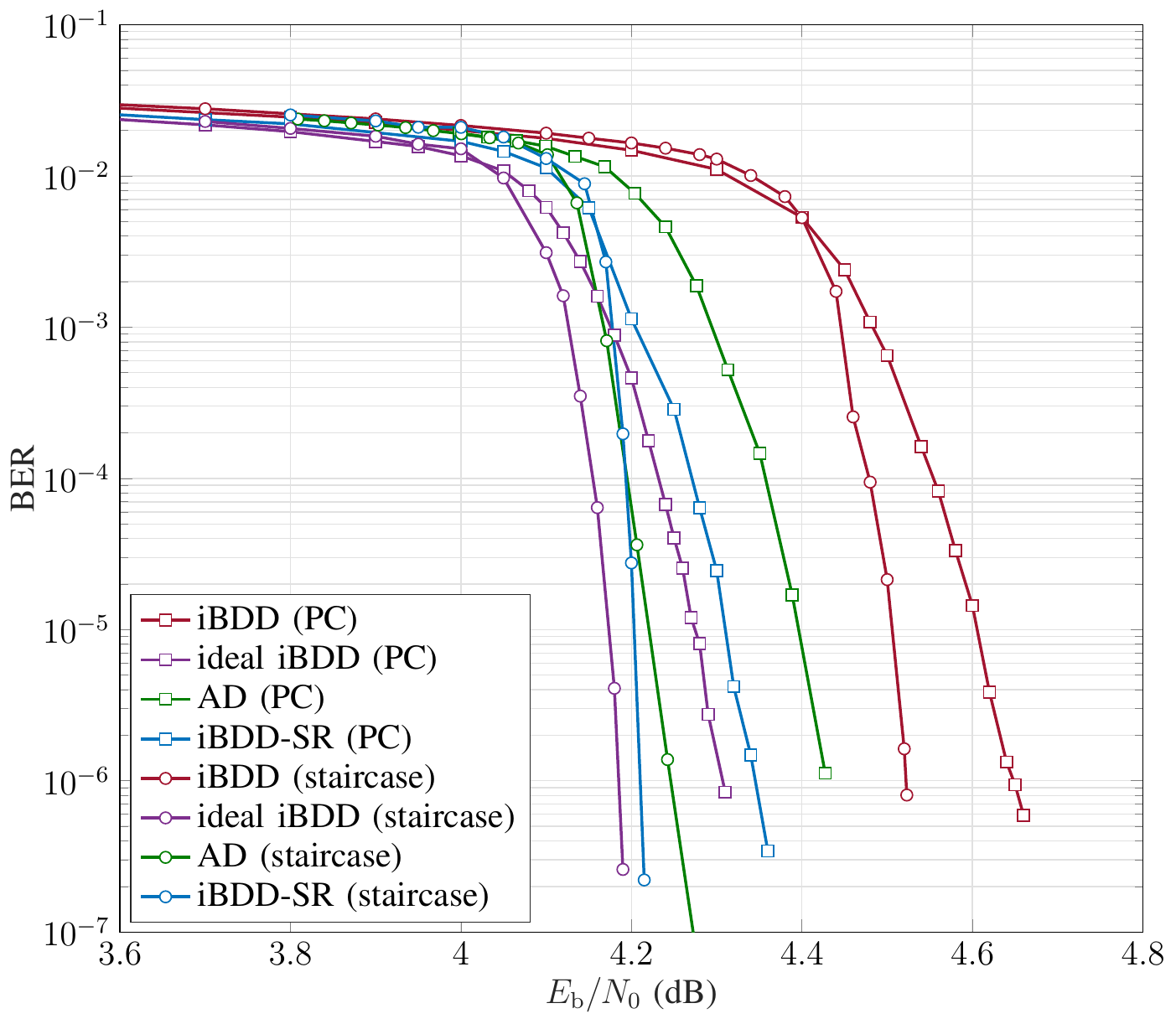}  
	\caption{Performance of iBDD, ideal iBDD, AD, and iBDD-SR for a PC with component code ($255$,$231$,$3$) and a staircase code with component code ($254$,$230$,$3$).}  
	\label{SimPCST} 
	\vspace{-2ex}
\end{figure} 

\begin{figure}[t] \centering 
	\includegraphics[width=\columnwidth]{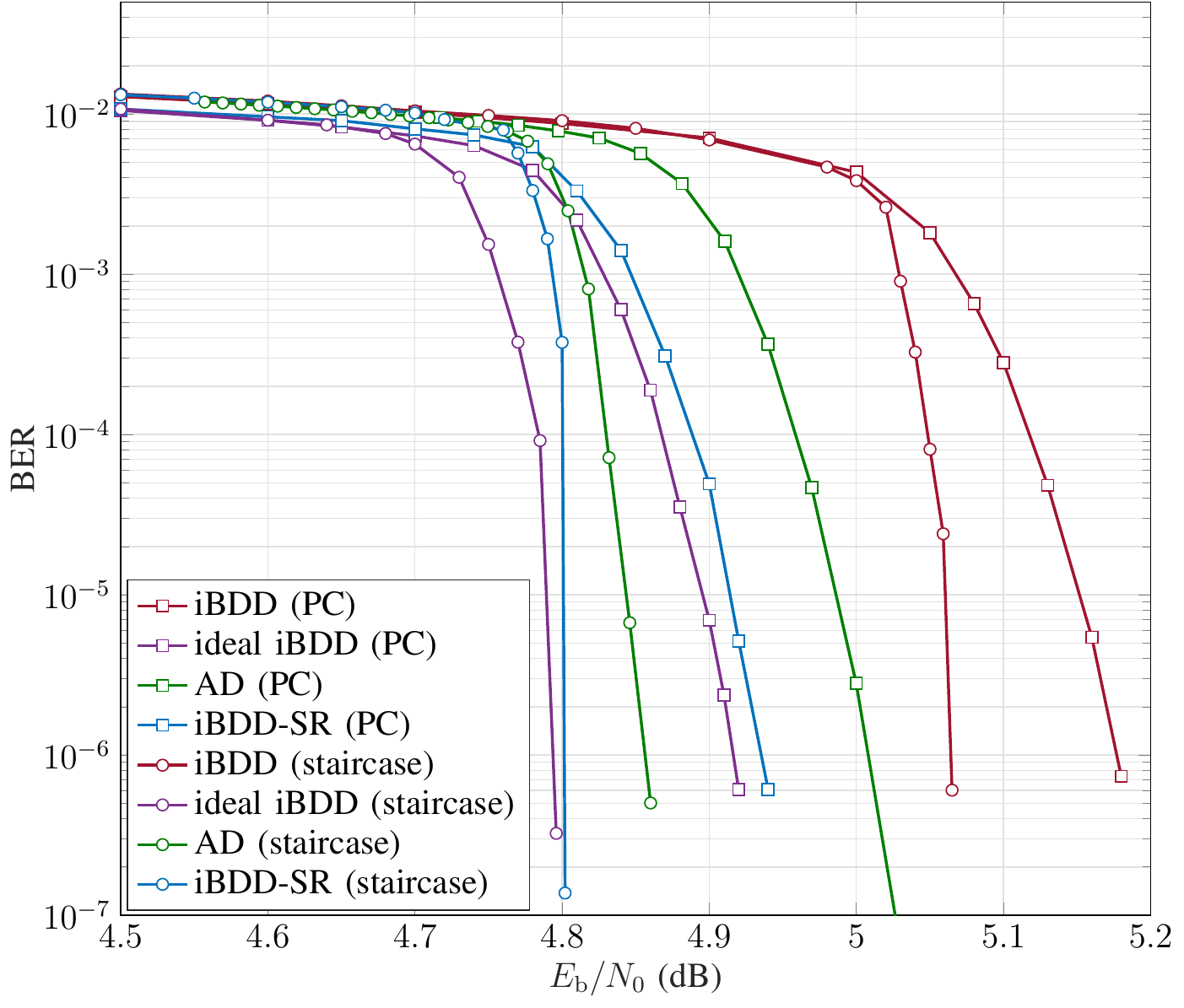}  
	\caption{Performance of iBDD, ideal iBDD, AD, and iBDD-SR for a PC with component code ($511$,$484$,$3$) and a staircase code with component code ($510$,$483$,$3$).}  
	\label{SimPCST1} 
	\vspace{-2ex}
\end{figure}

To evaluate the performance of iBDD-SR, we simulate the transmission of two PCs with ($255$,$231$,$3$) BCH component codes and ($511$,$484$,$3$) BCH component codes and two staircase codes with the same component codes shortened by one bit (i.e., ($254$,$230$,$3$) and ($510$,$483$,$3$)), over the bi-AWGN channel.\footnote{We are particularly interested in the performance of codes based on BCH component codes with $t=3$, since their decoders can be efficiently implemented via lookup tables \cite[Appendix I]{staircase_frank}.}  The code rate of the resulting product codes is  $R=0.820$ and $0.897$ for $n=255$ and $n=511$, respectively. The code rate of the staircase codes is
 $R=0.811$ and $0.894$ for $n=254$ and $n=510$, respectively. For the sake of comparison, we also simulate the performance of the codes under iBDD, ideal iBDD (i.e., a genie-aided iBDD where misscorrections are avoided by providing the component decoder input at its output whenever the error correction capability of the component code is exceeded), and AD. The BER performance of the considered codes is shown in Figs.~\ref{SimPCST}-\ref{SimPCST1}.

It is important to remark that if the channel LLRs are highly reliable but with wrong sign, one can expect that the decoding rule in \eqref{eq:GMDchrel_VN_scale}--\eqref{eq:BDDchrel_VN} will be unable to recover from these errors. In this situation, although ${\bar\mu}_{i,j}^{\mathsf r}$ may correspond to a correct decision, it is overridden by the channel, i.e., the hard decision on code bit $c_{i,j}$ made by the $i$-th row decoder, $\psi_{i,j}^{\mathsf{r},(\ell)}$, becomes  $\psi_{i,j}^{\mathsf{r},(\ell)}=\BB(w_i^{\mathsf r, (\ell)} \cdot \bar{\mu}_{i,j}^{\mathsf r,(\ell)} + L_{i,j})=\BB(L_{i,j})$ (cf. \eqref{eq:GMDchrel_VN_scale} and \eqref{eq:BDDchrel_VN}), which leads to an erroneously decoded bit. Therefore, one needs to be careful when applying iBDD-SR to avoid the appearance of an error floor. In particular, to avoid such errors and the presence of a high error floor, we run iBDD-SR for some iterations and then we append a few conventional iBDD iterations, where the channel reliabilities are disregarded when making the decision on a given code bit. The appended iBDD iterations increase the chance to correct transmission errors with high channel reliability. By doing so, an error floor is avoided. For PCs, we consider a maximum of $10$ iBDD-SR iterations followed by $2$ conventional iBDD iterations. For a fair comparison, for iBDD, ideal iBDD, and AD, we use a maximum of $12$ iterations. For staircase codes, we use a window decoder with window size of $7$ blocks and a maximum of $10$ iBDD-SR iterations followed by $2$ conventional iBDD iterations. 


Table~\ref{Tabcomp} summarizes the gains of iBDD-SR, ideal iBDD, and AD over conventional iBDD (fifth column)  and the gap to the corresponding capacity (sixth column) at the BER of $10^{-6}$ for both PCs and staircase codes with BCH component codes of parameters  ($255$,$231$,$3$) and ($254$,$230$,$3$), respectively. Whether the decoder exploits the channel reliabilities and the nature of the messages exchanged in the iterative decoding (hard or soft) is indicated in the third and fourth column, respectively. iBDD-SR yields a performance gain of $0.29$ dB and $0.31$ dB with respect to iBDD for the PC and the staircase code, respectively. For very high-throughput applications such as fiber-optic communications such gains are significant. Furthermore, one can see that iBDD-SR performs  close to the ideal iBDD for both PCs and staircase codes. Interestingly, the staircase code with iBDD-SR outperforms the PC with ideal iBDD by $0.1$ dB at a BER of $10^{-6}$. 

In Table~\ref{Tabcomp1}, we give the gains of iBDD-SR, ideal iBDD, and AD over conventional iBDD  and the gap to the corresponding capacity at the BER of $10^{-6}$ for both PCs and staircase codes with BCH component codes of parameters ($511$,$484$,$3$) and ($510$,$483$,$3$), respectively. Compared to the shorter block length, one can see that iBDD-SR yields similar gains with respect to iBDD, albeit slightly smaller. Also, the gap between iBDD-SR and ideal iBDD reduces even further. In particular, for the staircase code iBDD-SR performs almost the same as miscorrection-free decoder. Furthermore, the gap to capacity for all decoders is also reduced.

From the GN model, one can see that an optical link SNR improvement of
$a$ dB  yields $a$ dB of optical reach enhancement \cite[Eq.~59]{Poggiolini_applica}. Thus, the $~0.3$ dB performance improvement of iBDD-SR over iBDD yields about $7.2\%$ of reach enhancement. We also remark that for extended BCH component codes with $t=2$, not reported here, the gain of iBDD-SR with respect to iBDD is slightly larger and iBDD-SR performs almost identical as conventional iBDD (see Fig.~3 in \cite{She18b1} for a curve).

\begin{figure}[t] \centering 
	\includegraphics[width=\columnwidth]{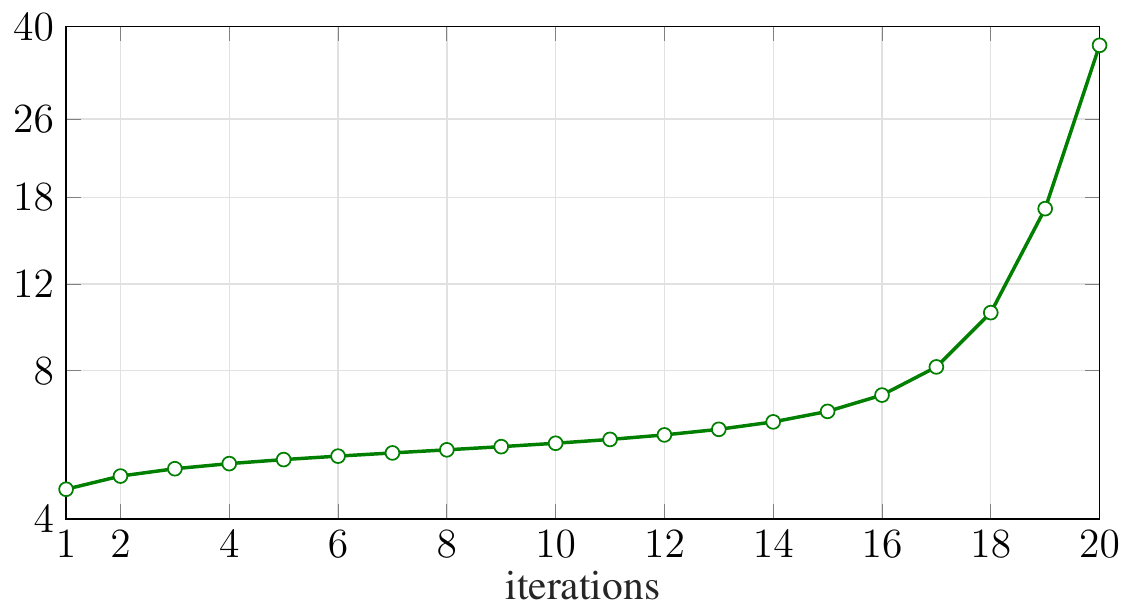}  
	\caption{The evolution of scaling factors for the GLDPC code ensemble with BCH component code ($255$,$231$,$3$) over $20$ iterations.}  
	\label{wPC} 
	\vspace{-2ex}
\end{figure} 

\begin{figure}[t] \centering 
	\includegraphics[width=\columnwidth]{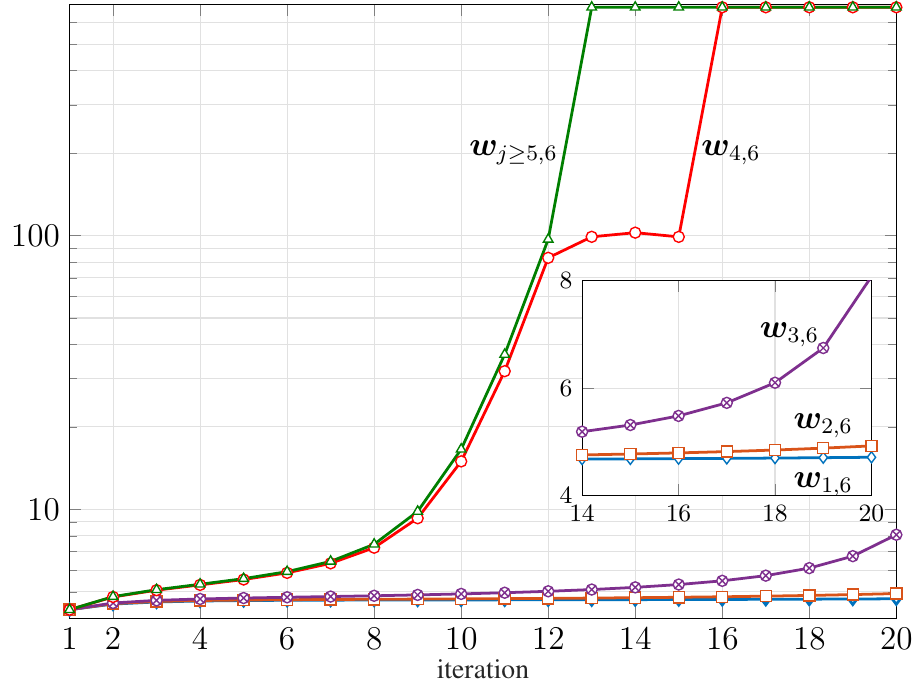}  
	\caption{The evolution of scaling factors for the SC-GLDPC code ensemble with BCH component code ($255$,$231$,$3$) over $20$ iterations, using a window decoder of window size $6$, for the $6$-th spatial position within the window.}  
	\label{wSC} 
	\vspace{-2ex}
\end{figure} 
In Fig.~\ref{wPC}, we show the evolution of  the scaling factors resulting from the density evolution for the GLDPC code ensemble with BCH component code ($255$,$231$,$3$) as a function of the number of \emph{half} iterations, where each iteration corresponds to one row or column decoding step. The scaling factors in Fig.~\ref{wPC} are obtained at the decoding threshold, which is $4.18$ dB. As can be seen, the scaling factors are monotonically increasing over iterations. This is expected, as the reliability of the BDD output increases with the number of iterations. In Fig.~\ref{wSC}, we show the evolution of the scaling factors for the SC-GLDPC code ensemble with the same component code. In the figure, $\ww_{a,j}$ corresponds to the vector of scaling factors corresponding to the $j$-th position inside the decoding window (of size $U$) containing  spatial positions $[a,\ldots,a+U-1]$. In particular, the figure plots the scaling factors obtained by the density evolution (at the code threshold, i.e., $4.05$ dB)
for the $6$-th spatial position inside the windows. Note that after $5$ ($U-1$) window slides, the scaling factors converge to a given value. The same phenomenon has been observed for other spatial positions within the window and other ensembles. Thus, from a practical viewpoint, it is not necessary to store the scaling factors for every spatial position, but only for a limited number of positions.   

It is important to remark that the density evolution in Section~\ref{sec:DE_PCs} assumes the exchange of extrinsic information between component decoders. However, \emph{extrinsic} message passing decoding of product-like codes (as explained in Section~\ref{sec:DE_PCs} and \cite{JianPfister2017} for iBDD-SR and iBDD, respectively) is complex, as it requires that the component codes are decoded $n$ times in each iteration. In practice, when very high throughputs are required, product-like codes are hence decoded using conventional iterative row/column decoding of the component codes, i.e., \emph{intrinsic} message passing, and this is the algorithm that we used for the simulations. Thus, the scaling factors are obtained for a slightly different decoder than the one used in practice. One may then wonder how good the scaling factors derived for the extrinsic message passing are for the conventional decoder. To verify the impact of using the scaling factors found from the density evolution for finite-length codes with conventional row/column decoding, we also performed a search based on Monte-Carlo simulations to optimize the scaling factors for PCs,  with a grid search of step size of $0.01$. Interestingly, the performance of  iBDD-SR with scaling factors from the density evolution yields slightly better results (probably due to the non-exhaustive limited Monte-Carlo search), supporting that their derivation using the density evolution is very useful in practice. Also, we would like to stress that the Monte-Carlo search is not feasible for staircase codes, thus the derived density evolution is crucial in the design of iBDD-SR in this case.

%
%
%

\section{Conclusion}   

We proposed iterative bounded distance decoding with scaled reliability, a new decoding algorithm for the decoding of product-like codes. The proposed algorithm, based on BDD of the component codes, exploits the channel reliabilities but, notably, is a binary message passing algorithm, i.e., the component decoders exchange only hard decisions. The proposed algorithm improves the performance of conventional iBDD, with the same decoder data flow, at the expense of a minor increase in complexity. For two particular PCs and staircase codes built from BCH component code codes $(255,231,3$) and $(511,484,3)$ (shortened by one bit for staircase codes), the proposed algorithm outperforms conventional iBDD by $0.24$--$0.31$ dB and yields performance very close to that of ideal iBDD without miscorrections. For a PC with $(255,231,3)$ BCH component codes, in \cite{FougstedtiBDDSR2018}  we implemented iBDD-SR   in a 28nm process technology, achieving 1 Tb/s with an area and energy dissipation less than half of that of staircase decoders, at similar estimated net coding gains.

The proposed algorithm is appealing for applications requiring very  high throughputs such as fiber-optic communications.

\section*{Appendix}
\label{app:AppendixA}

We clarify the derivation of ${\Pue}\left( i \right)$. Recall that ${\Pue}\left( i \right)$ corresponds to the probability that BDD results in a codeword
(within Hamming distance $t$ of the input vector, see \eqref{eq:BDD_VN})  that has an error in the randomly selected position, given that the input vector has an error in that position and contains $i$ errors in the other $n-1$ positions.  One can easily see that if $0 \le i \le t - 1$, the total number of errors is less than or equal to $t$, meaning that the decoder is able to correct the codeword, therefore ${\Pue}\left( i \right)=0$. On the other hand, if $n-t-1 \le i \le n - 1$, the Hamming distance between the input vector and the all-one codeword is less than or equal to $t$, therefore  BDD decodes onto the all-one codeword and the randomly selected position will always be in error, i.e., ${\Pue}\left( i \right)=1$. 

The nontrivial case corresponds to $t\le i \le n-t-2$. In this case, we need to compute the probability that there exists a codeword $\bm c$ that has a one in the randomly-chosen position and is within Hamming distance $t$ of the weight-$(i+1)$  vector at the input of the bounded distance decoder, which we denote by $\bm r$. To do so, we can exploit the weight enumerator of the BCH code, $A_h$, and consider the equivalent problem of computing the probability that, given a codeword $\bm c$ of a given weight, the randomly selected bit for the input vector corresponds to an entry where $\bm c$ is one and the $i$ other ones of the input vector are placed such that $\ham(\bm c, \bm r)\le t$. 

We proceed as follows. 
Consider codewords of weight $h+1$, the total number of which is $A_{h+1}$. 
For a given codeword of weight $h+1$, the probability that the randomly selected erroneous bit is chosen among the codeword bit positions that are one is $\frac{{{h}+1}}{n}$.
Now, assume that the input vector has $h-j$ ones in $h-j$ out of the $h$ entries (one entry is already fixed) where the given codeword has ones. Thus, the input vector has $i-(h-j)$ ones in $i-(h-j)$ out of the $n-h-1$ entries where the given codeword is zero. The number of  possibilities is
\begin{align*}
\Big( {\begin{array}{*{20}{c}}
		{h}\\
		{h - j}
		\end{array}} \Big)
\end{align*}
 and 
\begin{align*}		
		\Big( {\begin{array}{*{20}{c}}
		{n - h - 1}\\
		{i-(h-j)}
		\end{array}} \Big)=\Big( {\begin{array}{*{20}{c}}
		{n - h - 1}\\
		{\delta  - j}
		\end{array}} \Big),
\end{align*}
respectively, where we defined $\delta\triangleq i-h+2j$ for convenience. Thus, the probability that this occurs is 
\begin{align*}
\frac{
\Big( {\begin{array}{*{20}{c}}
		{h}\\
		{h - j}
		\end{array}} \Big)\Big( {\begin{array}{*{20}{c}}
		{n - h - 1}\\
		{\delta  - j}
		\end{array}} \Big)}
		{\Big( {\begin{array}{*{20}{c}}
		{n - 1}\\
		i
		\end{array}} \Big)},
\end{align*}		
which we defined as $F^{(1)}_{h}$ in \eqref{f_pe}.

Finally, we need to sum over all cases such that $\bm r$ and the candidate codeword $\bm c$ are within Hamming distance $t$. Note that $\ham(\bm c,\bm r)=j+(i-(h-j))=i-h+2j=\delta$ and we need that $\delta\le t$. Thus, we need to sum over $\delta=1,\ldots,t$ and subsequently $j=0,\ldots,\delta$, which  results in the expression in \eqref{Pe}.

\section*{Acknowledgment}

The authors would like to thank Dr. Christian H\"ager and Prof. Henry Pfister for  fruitful discussions and providing the simulation results of AD  in Figs.~\ref{SimPCST}--\ref{SimPCST1}. 

\bibliographystyle{IEEEtran}

\end{document}